\documentclass[aps,prd,11pt,showpacs,superscriptaddress,amssymb,amsmath,nofootinbib]{revtex4-1}

\usepackage{amsmath,amssymb}
\usepackage{graphicx}
\usepackage{xcolor}
\usepackage{booktabs}
\usepackage{hyperref}
\usepackage{placeins}

\newcommand{\safeincludegraphics}[2][]{%
\IfFileExists{#2}{\includegraphics[#1]{#2}}{\fbox{\parbox[c][0.22\textheight][c]{0.9\linewidth}{\centering Figure file not found: #2}}}%
}

\begin{document}

\title{ Interpreting quantum coherence of neutrinos in asymptotically flat
    spacetimes via effective distances: Solution to the paradoxes of the
    Leggett-Garg inequality and quantum information quantities }

\author{Shan Wu}\email{1020251508@glut.edu.cn}

\author{Shu-Jun Rong}
\email{rongshj@glut.edu.cn}

	\begin{abstract}
	In this work, we propose a notion called effective distance to characterize the essential gravitational effects on quantum coherence of neutrinos.  Employing effective distances, we prove that one can always map neutrino-oscillation phases in asymptotically flat spacetimes onto the phases in flat spacetime, which is considered as a nontrivial realization of equivalence principle. Hence, the maximal amount of quantum coherence, quantified by the Leggett-Garg (LG) parameter  is independent from background spacetimes, namely, the paradox that the gravitational effects on quantum coherence of neutrinos are  measure-dependent (Ettefaghi et al.~\cite{Ettefaghi:2022nsq}) does not arise. Following the same route, we preclude the gravitational amplifying or damping of amplitudes of the quantum information quantities for special masses of black holes (Wang et al.~\cite{Wang:2024tfh}).
To demonstrate furthermore the independence of quantum coherence from  background spacetimes, we examine the proposal using the LG parameter
to identify quantum-corrected models of spacetimes. For different emission positions of neutrinos and various quantum parameters,  we find no consistent patterns in oscillating LG curves to discriminate a quantum metric from the classical one.
	\end{abstract}

\maketitle

\section{Introduction}

Due to feeble interactions, neutrinos can preserve their quantum coherence over astrophysical distances. This unique property renders them not only natural probes for physics beyond the Standard Model but also a novel testing ground at the intersection of relativistic quantum information~\cite{Blasone:2007vw,Blasone:2007wp,Bittencourt:2023asd,Blasone:2014jea,Blasone:2015lya,Alok:2014gya,Banerjee:2015mha,Ettefaghi:2020otb,Banerjee:2025gau,Dixit:2019swl,Siwach:2022xhx,Dixit:2023fke,Konwar:2024pkh,Alok:2024xeg,SinghKoranga:2024rum,Nambiar:2026zhs,Yadav:2026lsx,Wang:2024tfh} and quantum gravity phenomenology~\cite{Stuttard:2020qfv,Hellmann:2022cgt,DEsposito:2023psn,Banerjee:2022slh}. In particular, when these messengers traverse curved spacetime---whether in the vicinity of black hole horizons, or in the presence of cosmological curvature---we would confront a question: how are their quantum  properties modified by gravitational effects? This question is not merely concerning propagations of realistic high-energy astrophysical neutrinos~\cite{IceCube:2013low,IceCube:2013cdw,IceCube:2014stg}; it also shows a subtle interplay between quantum mechanics and general relativity.

The Leggett-Garg inequality (LGI)~\cite{Leggett:1985} serves as the temporal analogue of Bell's inequality, providing a criterion for testing macrorealism in a system. A violation of the LGI signifies the presence of quantum coherence. In the context of neutrino oscillations, the flavor state of a neutrino is the coherent superpositions of mass eigenstates, and the flavor conversion probabilities directly reflect the evolution of the quantum coherence. Thus, the LGI offers a tool for quantifying the quantumness of neutrino oscillations. Recently, the LGI has been applied to neutrino oscillation data from experiments such as Daya Bay~\cite{Fu:2017hky}, MINOS~\cite{Formaggio:2016cuh}, and KamLAND~\cite{Wang:2022tnr}, yielding significant violations. These results not only confirm the quantum nature of neutrino oscillations but also demonstrate that neutrinos can maintain non-classical correlations over macroscopic distances.

However, the existing LGI studies on neutrinos are mostly conducted under the assumption of flat spacetime, thereby neglecting the possible modulations of quantum coherence by gravitational fields. As is known, the interplay between gravity and quantum coherence was investigated half a century ago by the Colella-Overhauser-Werner (COW) neutron interferometry experiment~\cite{Colella:1975COW}, whose outcomes show that the Newtonian gravity can work in quantum mechanics thorough the phases in wave functions. Based on the Newtonian quantum phases, tabletop experiments to
witness quantumness of gravity with gravity-induced entanglement of masses are proposed~\cite{Bose:2017nin,Marletto:2017kzi}, which shows that gravitational effects on quantum phases may provide alternative avenue to probe the nature of spacetime.

Gravity modifies the phase evolution of a quantum state through changes in the spacetime metric; moreover, the difference in proper time experienced by observers at different spatial positions can further lead to the degradation of quantum coherence~\cite{Chatelain:2019nkf,Petruzziello:2020wea,Luciano:2021gdp}. These considerations lead to a pivotal question: how can such effects be quantitatively revealed through the degree of LGI violation?
On this question, Ettefaghi et al.~\cite{Ettefaghi:2022nsq} investigated the quantum coherence of two-flavor neutrinos in the Schwarzschild spacetime, performing a quantitative analysis via the LG parameter $K_{3}$ and the $l_{1}$-norm coherence measure $\mathcal{C(\rho)}$. They found that for certain energy ranges, gravity suppresses the maximum of $K_{3}$, but keeps the maximum of $\mathcal{C(\rho)}$ invariant. This seemingly paradoxical result highlights a fundamental subtlety: different measures of quantumness respond to gravitational backgrounds in qualitatively distinct ways.
Yet we should note that their observation is based on the measurement scheme of $K_{3}$ where uniform proper-distance intervals are taken. Considering the essential gravitation effects on neutrino phases, the obtained $K_{3,\max}$ may not be the genuine maximum.
Similarly, for the quantum information quantities which can be expressed by neutrino phases, we should also examine the paradox that gravity amplifies or damps the amplitudes of information quantities for special values of a metric parameter, e.g. Ref.\cite{Wang:2024tfh}.

In this wok, we aim to identify the essential gravitational effects on the quantum phases of neutrinos in a Hamiltonian formalism. To this end, we propose a notion called effective distance to separate the impact of gravity and that of the intrinsic properties of neutrinos on their quantum coherence.
Employing effective distances, we demonstrate that one can map the oscillation phase in an asymptotically flat spacetime  onto the phase in flat spacetime.
Consequently, the maximum of a measure for quantum coherence in curved spacetime can be kept the same as that in flat spacetime. For the LG parameter, using the measurement scheme where uniform effective-distance intervals are taken, we show that the genuine maximum takes the unique value determined by the properties of neutrinos themselves.

Besides the essential gravitational effects on quantum coherence of particles, modulations on entanglement phases are employed to explore properties of a spacetime.
In particular, the Bell's inequality is recently proposed to discriminate classical and quantum-corrected metrics~\cite{Petruzziello:2023xhb}.
Following the same route, we examine the proposal using the LG parameter to probe quantum-corrected models of spacetime.
We analyze the response of LG parameters to the classical and quantum-corrected Schwarzschild and Kerr metrics, with the measurement scheme choosing uniform proper-distance intervals.

The remainder  of the paper is organised as follows. We first present a theoretical framework to treat neutrino oscillations in curved spacetimes and define the LG parameter in Sec.~\ref{sec:theo}. We choose a Hamiltonian formalism  and introduce effective distances to determine the gravitational effects. In Sec.~\ref{sec:Sch resp}, we show the LG parameter with different measurement schemes and analyse the essential and apparent quantum coherence in the Schwarzschild spacetime.
In Sec.~\ref{sec:Compa Schw}, we compare the response of the LG parameter to the classical and quantum-corrected Schwarzschild metrics. In Sec.~\ref{sec:Compa Kerr}, the comparison of the LG response is performed for classical and quantum-corrected Kerr metrics.
We conclude in Sec.~\ref{sec:Conclu}. Throughout the paper, we take the units $G=\hbar=c=1$.

\section{Theoretical Framework}
\label{sec:theo}
\subsection{Three-Flavor Neutrino Oscillation}

In the standard three-flavor neutrino framework, the flavor eigenstates
\(|\nu_\alpha\rangle\) and the mass eigenstates \(|\nu_i\rangle\) are related by the
Pontecorvo--Maki--Nakagawa--Sakata (PMNS) mixing matrix \(U\)
~\cite{Maki:1962mu,Navas:2024}:
\begin{equation}
|\nu_\alpha\rangle=\sum_{i=1}^{3}U_{\alpha i}^{*}|\nu_i\rangle,
\qquad
|\nu_i\rangle=\sum_{\alpha=e,\mu,\tau}U_{\alpha i}|\nu_\alpha\rangle .
\end{equation}
Here \(\alpha=e,\mu,\tau\) and \(i=1,2,3\). The PMNS matrix is written in the
standard parametrization~\cite{Navas:2024} as
\begin{equation}
U=
\begin{pmatrix}
c_{12}c_{13} & s_{12}c_{13} & s_{13}e^{-i\delta_{\rm CP}}\\
-s_{12}c_{23}-c_{12}s_{13}s_{23}e^{i\delta_{\rm CP}} &
c_{12}c_{23}-s_{12}s_{13}s_{23}e^{i\delta_{\rm CP}} &
c_{13}s_{23}\\
s_{12}s_{23}-c_{12}s_{13}c_{23}e^{i\delta_{\rm CP}} &
-c_{12}s_{23}-s_{12}s_{13}c_{23}e^{i\delta_{\rm CP}} &
c_{13}c_{23}
\end{pmatrix},
\end{equation}
where \(s_{ij}=\sin\theta_{ij}\) and \(c_{ij}=\cos\theta_{ij}\).

After subtracting the term proportional to the identity matrix, which only
contributes an overall phase and does not affect oscillation probabilities, the
vacuum effective Hamiltonian in flavor space is~\cite{Sarkar:2020vob,Farzan:2017xzy}
\begin{equation}
H_{\rm vac}(E)
=
\frac{1}{2E}
U
\begin{pmatrix}
0 & 0 & 0\\
0 & \Delta m_{21}^{2} & 0\\
0 & 0 & \Delta m_{31}^{2}
\end{pmatrix}
U^\dagger .
\label{eq:Hvac}
\end{equation}
In flat spacetime, after a propagation distance \(L\), the evolution operator and
the oscillation probability are respectively~\cite{Rahaman:2021,Harrison:2002ee}
\begin{equation}
\label{eq:flat}
S(L)=\exp[-iH_{\rm vac}(E)L],
\end{equation}
\begin{equation}
P_{\alpha\rightarrow\beta}(L)
=
|\langle \nu_\beta|S(L)|\nu_\alpha\rangle|^2 .
\end{equation}

\subsection{ Neutrino Propagation in Curved Spacetime and Effective Distance}

The covariant propagation phase of a neutrino
mass eigenstate in curved spacetime can be derived from the Hamilton--Jacobi equation, as detailed
in Appendix~\ref{app:covariant-phase}. Using the  propagation phases in terms of the
locally measured energy \(E_{\rm loc}\) and the local propagation distance
\(dL_{\rm loc}\), the full flavor-evolution operator can be written as
\begin{equation}
S(L)
=
\exp\left[
-\frac{i}{2}
U
\begin{pmatrix}
0 & 0 & 0\\
0 & \Delta m_{21}^{2} & 0\\
0 & 0 & \Delta m_{31}^{2}
\end{pmatrix}
U^\dagger
\int_{0}^{L}
\frac{dL_{\rm loc}'}
{E_{\rm loc}\!\left[x(L_{\rm loc}')\right]}
\right].
\end{equation}
Here, the quantity
\(x(L_{\rm loc}')\) collectively denotes the position along the trajectory and
may include \(r(L_{\rm loc}')\), \(\theta(L_{\rm loc}')\), and any other
coordinates on which the local energy depends. For an asymptotically flat spacetime, the relation between  \(E_{\rm loc}\) and the energy at infinity
\(E_\infty\) is expressed as
\begin{equation}
\mathcal{F}[x(L_{\rm loc})]
=
\frac{E_\infty}
{E_{\rm loc}[x(L_{\rm loc})]},
\end{equation}
where $\mathcal{F}[x(L_{\rm loc})]$ is the red-shift factor.
We introduce an effective distance defined as
\begin{equation}
\Lambda(L)
\equiv
\int_0^{L}
\mathcal{F}[x(L_{\rm loc}')]\,
dL_{\rm loc}'.
\end{equation}
It then follows that
\begin{equation}
\int_0^{L}
\frac{dL_{\rm loc}'}
{E_{\rm loc}[x(L_{\rm loc}')]}
=
\frac{\Lambda(L)}{E_\infty},
\end{equation}
and the flavor-evolution operator can therefore be written in a unified form
\begin{equation}
\label{eq:general}
S(L)
=
\exp\left[
-\frac{i}{2E_\infty}
U
\begin{pmatrix}
0 & 0 & 0\\
0 & \Delta m_{21}^{2} & 0\\
0 & 0 & \Delta m_{31}^{2}
\end{pmatrix}
U^\dagger
\Lambda(L)
\right]
=
\exp\left[
-iH_{\rm vac}(E_\infty)\Lambda(L)
\right].
\end{equation}
It is worth noting that the gravitational modulation on neutrino oscillations is completely determined by the effective distance. Furthermore, given an effective distance,
one can always map an neutrino oscillation phase in asymptotically flat spacetimes onto the one in flat spacetime, which
can be viewed as a realization of equivalence principle and can lead to  nontrivial consequences in the observation of quantum coherence of neutrinos .

Now we apply the general flavor evolution operator to the Schwarzschild-~\cite{Cardall:1996cd,Fornengo:1996ef} and Kerr-type spacetimes~\cite{Ren:2010yf,Swami:2022xet}.
For a static spherically symmetric spacetime, the line element is
\begin{equation}
ds^2
=
-f(r)dt^2
+
\frac{dr^2}{f(r)}
+
r^2d\Omega^2 .
\label{eq:static_metric}
\end{equation}
The energy at infinity \(E_\infty\) and that
measured by a local static observer satisfies
\begin{equation}
E_{\rm loc}(r)=\frac{E_\infty}{\sqrt{f(r)}} .
\end{equation}
If the local energy at the emission point is \(E_{\rm emit}^{\rm loc}\), then
\begin{equation}
E_\infty=E_{\rm emit}^{\rm loc}\sqrt{f(r_{\rm emit})}.
\end{equation}
Using the proper distance \(L\), we obtain the effective
distance
\begin{equation}
\Lambda(L)
=
\int_0^L \sqrt{f[r(L')]}\,dL'.
\label{eq:Lambda}
\end{equation}
We can check that the evolution operator gives neutrino-propagation phase which is of a covariant form as in Ref.\cite{Fornengo:1996ef}.

For outward radial propagation in the equatorial plane, the trajectory satisfies
\begin{equation}
\frac{dr}{dL}=+\sqrt{f(r)} .
\end{equation}
The corresponding effective distance is of a simple form, i.e.,
\begin{equation}
\Lambda_{\rm rad}(L)
=\int_{r_{e}}^{r_{d}}  \, dr=r_{d}-r_{e},
\end{equation}
with $r_{d}$, $r_{e}$ being radial coordinate at the detection and at the emission location respectively.

For nonradial propagation on the equatorial plane,
the orbital equation satifies
\begin{equation}
\left|\frac{dr}{dL}\right|
=
\sqrt{f(r)}
\sqrt{1-\frac{f(r)b^2}{r^2}},
\end{equation}
where the impact parameter
\begin{equation}
b=\frac{L_z}{E_\infty},
\end{equation}
with $L{_z}$ being the angular momentum at infinity.
If the local emission angle is \(\psi\), we get the
relation~\cite{Ishihara:2016vdc}
\begin{equation}
\sin\psi
=
\frac{b\sqrt{f(r_{\rm emit})}}{r_{\rm emit}},
\qquad
b=
\frac{r_{\rm emit}\sin\psi}{\sqrt{f(r_{\rm emit})}} .
\end{equation}
The effective distance for nonradial propagation reads
\begin{equation}
\Lambda_{\rm non}(L)
=
\int_0^L \sqrt{f[r_{\rm non}(L')]}\,dL'.
\end{equation}

For the  Kerr spacetime, the line element in Boyer--Lindquist coordinates is written as~\cite{Kerr:1963ud,Boyer:1967}
\begin{align}
ds^2_{\rm Kerr}
=&
-\left(1-\frac{2Mr}{\Sigma}\right)dt^2
-\frac{4Mar\sin^2\theta}{\Sigma}\,dt\,d\phi
+\frac{\Sigma}{\Delta_{\rm K}}dr^2
+\Sigma d\theta^2
\nonumber\\
&+
\left(
r^2+a^2+\frac{2Ma^2r\sin^2\theta}{\Sigma}
\right)\sin^2\theta\,d\phi^2 ,
\end{align}
where
\begin{equation}
\Sigma=r^2+a^2\cos^2\theta,
\qquad
\Delta_{\rm K}=r^2-2Mr+a^2 .
\end{equation}
Here, \(M\) is the black-hole mass, and \(a=J/M\) is the angular momentum per mass. The metric contains a nonzero \(g_{t\phi}\)
component, so the local energy relation needs to include the frame-dragging effect.
For a zero-angular-momentum observer (ZAMO), the local energy can be written as~\cite{Cheng:2025KTN}
\begin{equation}
E_{\rm loc}
=
\frac{E_\infty-\omega L_z}{\alpha}
=
\frac{E_\infty(1-\omega b)}{\alpha},
\end{equation}
where
\begin{equation}
\omega=-\frac{g_{t\phi}}{g_{\phi\phi}},
\qquad
\alpha=
\sqrt{
\frac{g_{t\phi}^{2}-g_{tt}g_{\phi\phi}}{g_{\phi\phi}}
}.
\end{equation}
Here \(\omega\) is the frame-dragging angular velocity, and \(\alpha\) is the lapse
function.

Considering  propagations on the equatorial plane,
in the case \(b=0\), the trajectory equation is~\cite{Bini:2008vk}
\begin{equation}
\frac{dr}{dL}
=
+\frac{\sqrt{\Delta_{\rm K}}}{r}.
\end{equation}
The corresponding effective distance is
\begin{equation}
\Lambda_{\rm rot}^{b=0}(L)
=
\int_0^L
\alpha\bigl[r_{b=0}(L')\bigr]\,dL' .
\end{equation}
For propagation with \(b\neq0\), the trajectory equation is
\begin{equation}
\frac{dr}{dL}
=
+
\frac{\alpha(r)\sqrt{\mathcal{R}_b(r)}}
{r^2[1-\omega(r)b]},
\end{equation}
where
\begin{equation}
\mathcal{R}_b(r)
=
\left(r^2+a^2-ab\right)^2
-
\Delta_{\rm K}(b-a)^2.
\end{equation}
The relation between \(b\) and the local emission angle \(\psi\) is~\cite{Stuchlik:2018KdS}
\begin{equation}
b
=
\frac{
\sqrt{g_{\phi\phi}(r_{\rm emit})}\sin\psi
}{
\alpha(r_{\rm emit})
+
\omega(r_{\rm emit})
\sqrt{g_{\phi\phi}(r_{\rm emit})}\sin\psi
}.
\end{equation}
The corresponding effective distance is
\begin{equation}
\Lambda_{\rm rot}^{b\neq0}(L)
=
\int_0^L
\frac{
\alpha[r(L')]
}{
1-\omega[r(L')]b
}
\,dL'.
\end{equation}

For non-equatorial three-dimensional propagations, the ZAMO lapse function
\(\alpha\) and the frame-dragging angular velocity \(\omega\) depend on both
\(r\) and \(\theta\). The corresponding effective distance is
\begin{equation}
\Lambda_{\rm rot}(L)
=
\int_0^L
\frac{
\alpha[r(L'),\theta(L')]
}{
1-\omega[r(L'),\theta(L')]b
}
\,dL'.
\end{equation}

\subsection{Two-Time Correlation Functions and the LG Parameter}

The LG inequality is formulated in terms of two-time correlation
functions defined along a temporal sequence. For neutrino oscillations, the
propagation time and propagation distance can be related to each other in the
relativistic approximation, so the time parameter can be equivalently expressed by
the propagation distance.

We define the dichotomic observable \(Q\) as~\cite{Emary:2013wfl}
\begin{equation}
Q=
\begin{cases}
+1, & \nu_e,\\
-1, & \nu_\mu,\nu_\tau .
\end{cases}
\end{equation}
For the measurement points which are chosen to be equally spaced in the proper propagation
distance, we have
\begin{equation}
L_i=i\Delta L,
\qquad i=1,2,3,4 ,
\end{equation}
where \(\Delta L\) is the proper distance interval between adjacent
points. The corresponding effective distances are
\begin{equation}
\Lambda_i=\Lambda(L_i),
\qquad i=1,2,3,4 .
\end{equation}

The two-time correlation function is defined as~\cite{Emary:2013wfl,Gangopadhyay:2017}
\begin{equation}
C_{ij}
=
\langle Q(L_i)Q(L_j)\rangle .
\end{equation}
For the initial state \(\nu_e\), it can be written as~\cite{Gangopadhyay:2017}
\begin{equation}
C_{ij}
=
\sum_{\alpha,\beta=e,\mu,\tau}
q_\alpha q_\beta
P_{e\rightarrow \alpha}(\Lambda_i)
P_{\alpha\rightarrow \beta}(\Lambda_j-\Lambda_i),
\label{eq:Cij}
\end{equation}
where \(q_e=+1\) and \(q_\mu=q_\tau=-1\).

The LG parameter corresponding to the four measurement points is defined as~\cite{Leggett:1985,Emary:2013wfl,Gangopadhyay:2017,Formaggio:2016cuh}
\begin{equation}
K_4
=
C_{12}+C_{23}+C_{34}-C_{14}.
\label{eq:K4}
\end{equation}
For a classical system satisfying macrorealism and noninvasive measurability, the
 upper bound is
\begin{equation}
K_4\leq 2 .
\end{equation}
When \(K_4>2\), the LG inequality is violated, indicating that the
neutrino oscillation process possesses nonclassical temporal correlations.
Therefore, \(K_4>2\) is used as the criterion for identifying quantumness in neutrino oscillations, and the maximal quantumness is
characterized by the parameter \(K_{4,\max}\).

For a given emission position \(x\), we define
\begin{equation}
K_{4,\max}(x)
=
\max_{\Delta L}K_4(\Delta L;x).
\label{eq:K4max}
\end{equation}
Here \(x\) is the dimensionless emission position,
\begin{equation}
x=\frac{r_{\rm emit}}{r_h},
\label{eq:x_emit}
\end{equation}
where \(r_{\rm emit}\) denotes the radial coordinate of the neutrino emission-point,
and \(r_h\) is the event-horizon radius of the corresponding spacetime.

\section{LG Response in Schwarzschild Spacetime }
\label{sec:Sch resp}
\subsection{\(K_{4,\max}\) in Flat and Schwarzschild Spacetime with Uniform Proper-distance Intervals}

In this work, the term 'response' refers to the variation of
\(K_{4,\max}(x)\) with the background metric, emission position, propagation mode. In all calculations, the initial state is taken
to be an electron neutrino \(|\nu_e\rangle\). Unless otherwise stated, all figures
below are obtained using the same neutrino oscillation parameters, local emission
energy, and black-hole mass. Specifically, the parameters are
\(\Delta m_{21}^{2}=7.50\times10^{-5}\ {\rm eV}^{2}\),
\(\Delta m_{31}^{2}=2.457\times10^{-3}\ {\rm eV}^{2}\),
\(\theta_{12}=33.48^\circ\), \(\theta_{23}=42.3^\circ\),
\(\theta_{13}=8.5^\circ\), and
\(\delta_{\rm CP}=306^\circ\)~\cite{Gonzalez-Garcia:2014bfa}.
The local emission energy is fixed at
\(E_{\rm emit}^{\rm loc}=1.5\times10^{12}\ {\rm eV}\), and the black-hole mass is taken to be
\(M=3.0\times10^{7}\ {\rm km}
\simeq 2.03\times10^{7}M_\odot\).
For radial propagation, we take \(b=0\), while for nonradial propagation the local
emission angle is chosen as \(\psi=75^\circ\). The
proper distance interval is taken in the range
\(\Delta L\in[1.0\times10^{8},\,5.0\times10^{8}]\ {\rm km}\).

\begin{figure*}[!t]
\centering
\safeincludegraphics[width=0.86\textwidth]{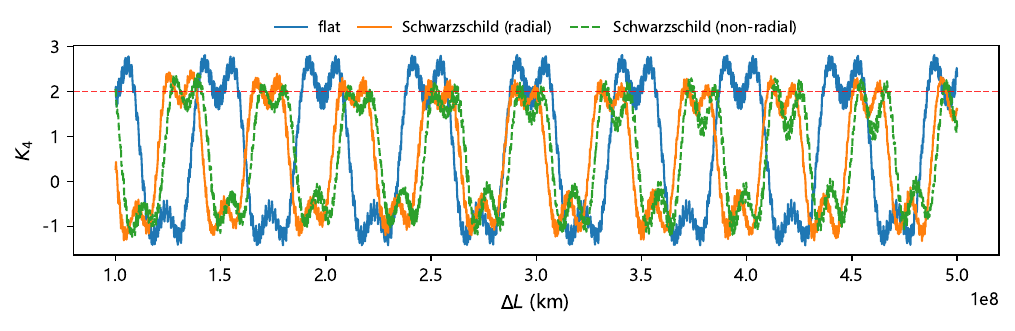}
\caption{
Variation of the LG parameter \(K_4\) with the proper interval
\(\Delta L\). The red dashed line denotes the classical LG upper bound
\(K_4=2\).
}
\label{fig:sch_k4}
\end{figure*}

To compare the maximal LG parameter in neutrino oscillations under
flat and Schwarzschild space-time, we plot \(K_4\) as a function
of \(\Delta L\) in Fig.~\ref{fig:sch_k4}. We can see that the three cases considered here all exceed the
classical upper bound  in part of the scanned interval, indicating the quantumness in neutrino oscillations. For the present
parameters, the maximum value in flat spacetime is
\(K_{4,\max}^{\rm flat}=2.809\), with the corresponding propagation interval
\(\Delta L_{\max}^{\rm flat}\simeq2.909\times10^8\,{\rm km}\).
For Schwarzschild radial and nonradial propagation, the maximum values are
\(K_{4,\max}^{\rm rad}\simeq2.466\) and
\(K_{4,\max}^{\rm non}\simeq2.396\), respectively, with the corresponding
propagation intervals
\(\Delta L_{\max}^{\rm rad}\simeq1.355\times10^8\,{\rm km}\) and
\(\Delta L_{\max}^{\rm non}\simeq1.390\times10^8\,{\rm km}\), respectively.

Compared with flat spacetime, the  gravitational field changes the
position and decrease the magnitude of \(K_{4,\max}\), thereby modifying
the observed quantumness in neutrino oscillations. This observation is consistent with the result obtained in Ref.~\cite{Ettefaghi:2022nsq}, which states that the maximum quantum coherence of neutrinos quantified by LG parameter is dependent on the spacetime background.
However, we should note that the observed quantumness is not essential since we choose a special measurement scheme with uniform proper-distance intervals. Choosing an alternative scheme with uniform effective-distance intervals,
we can show that \(K_{4,\max}\) in curved spacetimes recover the value in flat spacetime.

\subsection{Effective-Distance Interval Nonuniformity and the Essential Quantumness in Neutrino Oscillations }

\begin{figure*}[!t]
\centering
\safeincludegraphics[width=0.86\textwidth]{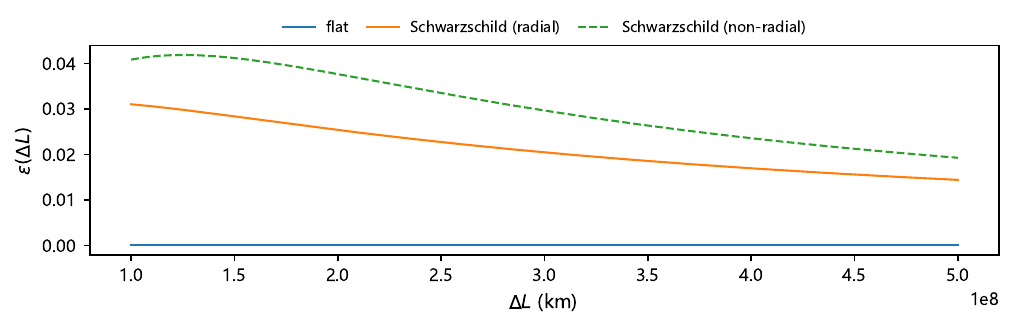}
\caption{
Variation of the effective-distance interval nonuniformity \(\epsilon(\Delta L)\)
with the proper interval \(\Delta L\). In flat spacetime, \(\epsilon=0\), whereas the
nonzero \(\epsilon\) in Schwarzschild spacetime indicates a nonuniform deformation
of the effective-distance intervals among the four measurement points.
}
\label{fig:sch_epsilon}
\end{figure*}

To understand the gravitational effects on quantum coherence in terms of effective distances, we introduce the interval
nonuniformity, defined as
\begin{equation}
\epsilon(\Delta L)
=
\frac{
\sqrt{
(d_{12}-\bar d)^2
+
(d_{23}-\bar d)^2
+
(d_{34}-\bar d)^2
}
}{\bar d}.
\label{eq:epsilon}
\end{equation}
Here,
\begin{align}
d_{12}
&=
\Lambda(2\Delta L)-\Lambda(\Delta L),
\nonumber\\
d_{23}
&=
\Lambda(3\Delta L)-\Lambda(2\Delta L),
\nonumber\\
d_{34}
&=
\Lambda(4\Delta L)-\Lambda(3\Delta L),
\nonumber\\
\bar d
&=
\frac{d_{12}+d_{23}+d_{34}}{3}.
\end{align}
The effective-distance-interval nonuniformity at the peak position is defined as
\begin{equation}
\epsilon_{\max}(x)
\equiv
\epsilon\bigl(\Delta L_{\max}(x);x\bigr),
\end{equation}
where
\begin{equation}
\Delta L_{\max}(x)
=
\operatorname*{arg\,max}_{\Delta L}K_4(\Delta L;x).
\end{equation}
According to this definition, we plot \(\epsilon(\Delta L)\) as a function of
\(\Delta L\). As shown in Fig.~\ref{fig:sch_epsilon},
\noindent
\(\epsilon=0\) is a benchmark in flat spacetime, while both radial and nonradial
propagation take a small but nonzero value, which violates the condition for the genuine maximum of the LG parameter. For the present parameters, when \(K_4\)
reaches its first maximum, the corresponding nonuniformities are
\(\epsilon_{\rm rad}(K_{4,\max})\simeq2.92\times10^{-2}\) and
\(\epsilon_{\rm non}(K_{4,\max})\simeq4.16\times10^{-2}\).
Although the nonuniformity is of order $10^{-2}$, the damping of \(K_{4,\max}\)
is pronounced in the gravitational field.

\begin{figure*}[!t]
\centering
\safeincludegraphics[width=0.86\textwidth]{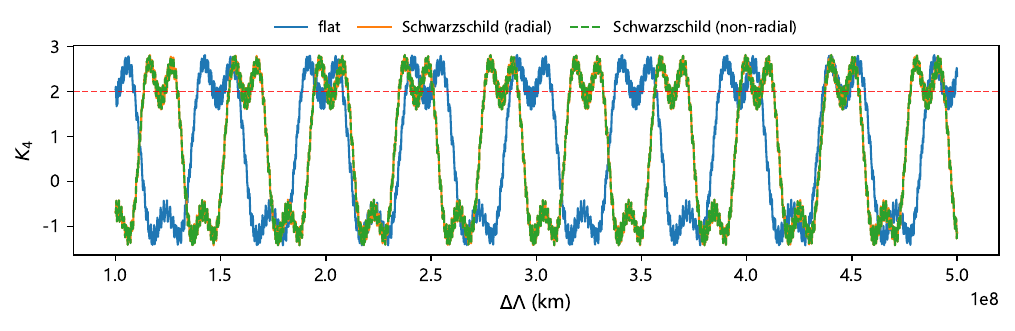}
\caption{
Variation of the LG parameter \(K_4\) with the uniform  effective-distance intervals among the four measurement points. For the Schwarzschild spacetime we take \(E_{\rm emit}^{\rm loc}=1.5\times10^{12}\ {\rm eV}\). For flat spacetime the energy $E$ also takes this value but the effective distance
satisfies  $\Lambda(L)=L$.
The phase shift between the flat and Schwarzschild case arises from the inequality $ E^{\rm loc}_{\rm emit}\neq E_{\infty}$.
\label{fig:effective1}
}
\end{figure*}

As we know, the quantumness shown in \(K_{4,\max}\) is dependent on the measurement scheme. Let us consider the case in which the effective-distance intervals are taken uniformly as in flat spacetime, namely
\begin{equation}
\Lambda(L_{4})-\Lambda(L_{3})=
\Lambda(L_{3})-\Lambda(L_{2})=
\Lambda(L_{2})-\Lambda(L_{1})=\Delta\Lambda=\Lambda(L_{1}).
\end{equation}
According to Eq. \ref{eq:general} and Eq. \ref{eq:Cij}, the gravitational effects on neutrino-propagation phases are just mapping the proper
distance to the effective distance, which can not change the flavor oscillation amplitudes.  Hence the genuine \(K_{4,\max}\) takes the same value as that in the flat sapcetime, see Fig.~\ref{fig:effective1}.
\noindent
 In fact, this value is determined by the PMNS matrix and the mass-squared differences of neutrinos.  Therefore, we can preclude the paradox that the maximum amount of quantum coherence in neutrino oscillations is damped by the gravitational field when quantified by the LG parameter, but kept invariant when quantified by $l_{1}$-norm measure~\cite{Ettefaghi:2022nsq}. It is worth noting that this observation also applies to gravitational modulations on other quantum information
quantities in terms of neutrino phases. Let us show this in the following section.

\subsection{The Essential Gravitational Effects on  Quantum-Information Quantities}
Quantum information quantities characterize properties of a quantum state, such as
coherence, quantum correlations, entanglement.
In flat spacetime, a general quantum-information quantity for 3-flavor neutrinos can be expressed as a
function of the oscillation phase \cite{Wang:2024tfh}:
\begin{equation}
\mathcal{Q}_{\rm flat}(L)
=
\mathcal{F}_{\mathcal{Q}}
\left[
\exp[i\Phi_{21}^{\rm flat}(L)],
\exp[i\Phi_{31}^{\rm flat}(L)]
\right].
\end{equation}
Here, $\mathcal{Q}$ represents an information quantity  with  \(\Phi_{ij}\) being the oscillation phase difference. The
variation of the quantum information quantity with propagation distance is determined by the two independent mass-squared differences.

In asymptotically flat spacetimes, gravity enters the oscillation
phase through the effective distance \(\Lambda(L)\):
\begin{equation}
\Phi_{ij}^{\rm grav}(L)
=
\frac{\Delta m_{ij}^{2}}{2E_\infty}
\Lambda(L).
\end{equation}
Correspondingly, quantum-information quantities can be written in a unified form
\begin{equation}
\mathcal{Q}_{\rm grav}(L)
=
\mathcal{F}_{\mathcal{Q}}
\left[
\exp[i\frac{\Delta m_{21}^{2}}{2E_\infty}\Lambda(L)],
\exp[i\frac{\Delta m_{31}^{2}}{2E_\infty}\Lambda(L)]
\right]
.
\label{eq:Sgrav}
\end{equation}
We can see that gravity induces a phase shift in
the oscillatory behavior of the quantum-information quantities as functions of
proper distance, without enhancing or suppressing their maximum values, namely, one can always map a value of a information quantity in curved spacetime onto the one in flat spacetime.
Hence, similar to the observation in the apparent LG paradox, several reported amplifying and damping of the amplitudes of quantum-information quantities of neutrinos by a special metric parameter such as mass of a black hole, e.g. Ref.~\cite{Wang:2024tfh},  can also be precluded.

\section{Comparison of the LG Parameters in Classical and Quantum-Corrected Schwarzschild Metrics}
\label{sec:Compa Schw}

We have shown that the essential maximal quantumness in neutrino oscillations is independent from sapcetime backgrounds satisfying the asymptotically flat condition. However, the gravitational modulations on the observed or apparent quantumness may still be used as a probe into nature of spacetime.
In particular, we note that the the Bell's inequality is recently proposed to discriminate classical and  quantum-corrected  metrics~\cite{Petruzziello:2023xhb}.  Following the proposal, in this section we examine whether the LG paratere can be employed to probe quantum-corrected models of spacetime.
We investigate the effects of quantum-corrected Schwarzschild metrics on the LG parameter with uniform proper-distance intervals. Hence the obtained \(K_{4,\max}\) is apparent instead of essential.

The first model considered here is the KS-type quantum-corrected metric\cite{Kazakov:1993ha,Konoplya:2019xmn}, with
\begin{equation}
f_{\rm KS}(r)
=
\frac{\sqrt{r^2-a^2}}{r}
-
\frac{2M}{r}
=
\sqrt{1-\frac{a^2}{r^2}}
-
\frac{2M}{r}.
\label{eq:f_KS}
\end{equation}
Here \(a\) is the quantum-correction parameter. When \(a=0\), this metric reduces
to the Schwarzschild metric. In the following calculation, we use the dimensionless
parameter \(a/r_h\), where \(r_h\) is the event-horizon radius of the KS spacetime.

The second model is the QCBH-type  metric
~\cite{Lewandowski:2022zce,Ali:2024ssf}, with
\begin{equation}
f_{\rm QCBH}(r)
=
1-\frac{2M}{r}
+
\frac{\alpha M^2}{r^4}.
\label{eq:f_QCBH_alpha}
\end{equation}
Introducing the dimensionless parameter
\begin{equation}
\bar{\alpha}=\frac{\alpha}{M^2},
\end{equation}
it can be written as
\begin{equation}
f_{\rm QCBH}(r)
=
1-\frac{2M}{r}
+
\bar{\alpha}\frac{M^4}{r^4}.
\label{eq:f_QCBH}
\end{equation}

\subsection{Response Comparison at the Fixed Emission Position \(x=3\)}

\begin{figure*}[!t]
\centering
\safeincludegraphics[width=0.86\textwidth]{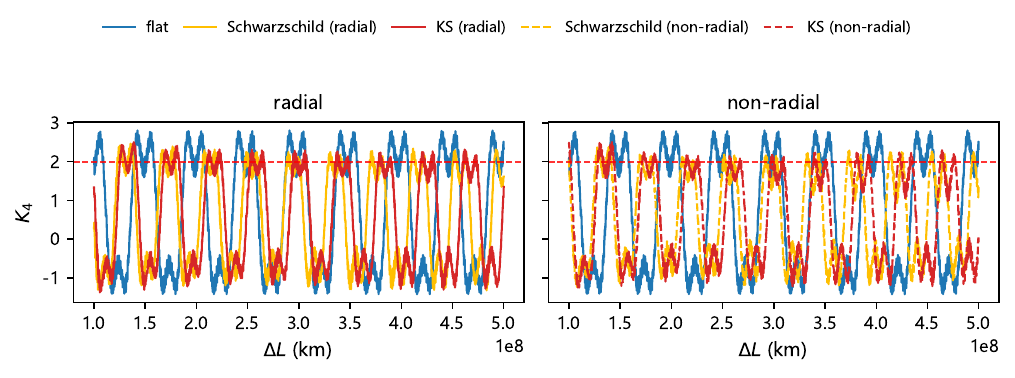}
\par\vspace{0.35em}
\vspace{0.5em}
\safeincludegraphics[width=0.86\textwidth]{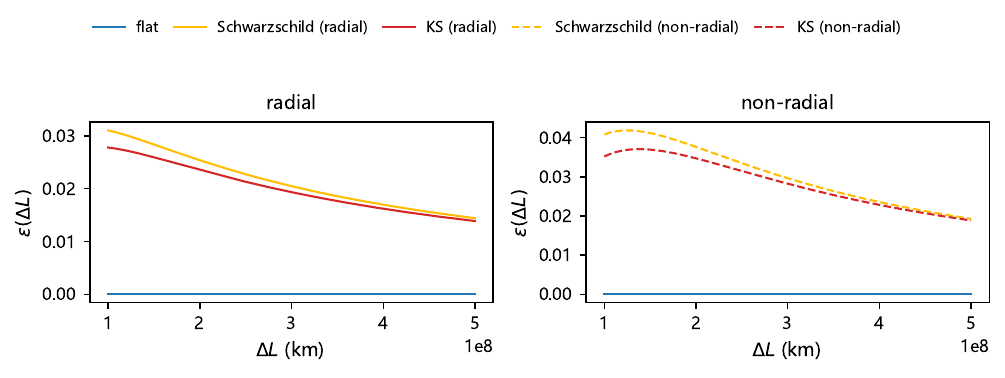}
\par\vspace{0.35em}
\caption{
 The variation of \(K_4(\Delta L)\) and \(\epsilon(\Delta L)\) with \(\Delta L\) in the
classical Schwarzschild and the KS quantum-corrected metric, at the fixed emission position \(x=r_{\rm emit}/r_h=3\). The upper panel shows
the LG parameter \(K_4(\Delta L)\), and the lower panel  shows the corresponding
effective-distance-interval nonuniformity \(\epsilon(\Delta L)\). In each panel, the left and
right subpanels correspond to radial and nonradial propagation, respectively.
}
\label{fig:ks_fixedx}
\end{figure*}

\begin{figure*}[!t]
\centering
\safeincludegraphics[width=0.86\textwidth]{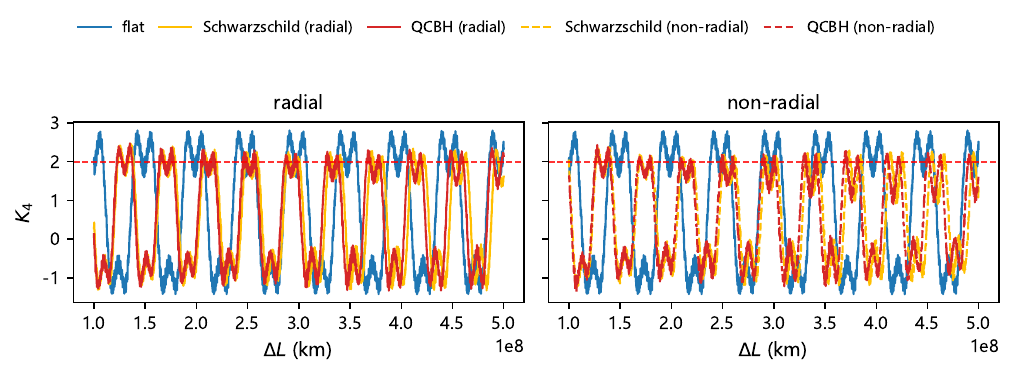}
\par\vspace{0.35em}
\vspace{0.5em}
\safeincludegraphics[width=0.86\textwidth]{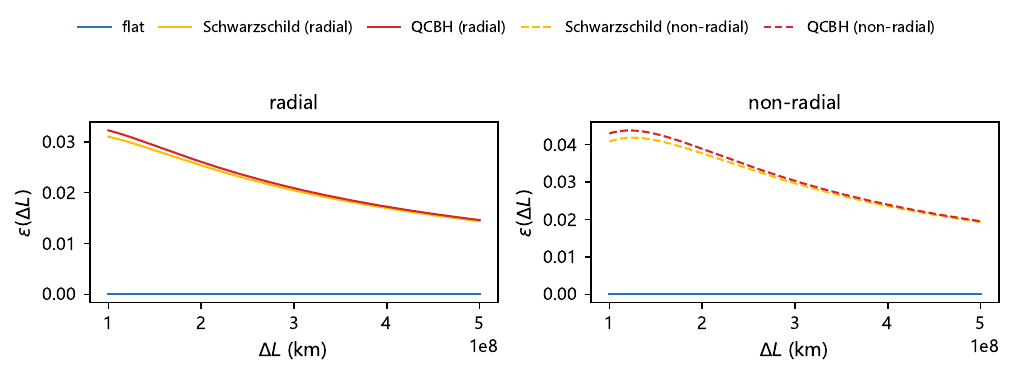}
\par\vspace{0.35em}
\caption{
The variation of \(K_4(\Delta L)\) and \(\epsilon(\Delta L)\) with \(\Delta L\) in the
classical Schwarzschild and the QCBH  quantum-corrected metric, at the fixed emission position \(x=r_{\rm emit}/r_h=3\).
}
\label{fig:qcbh_fixedx}
\end{figure*}

We first fix the emission position at \(x=r_{\rm emit}/r_h=3\) and compare \(K_{4,\max}\) in the
Schwarzschild, the KS, and the QCBH
metric. For the KS metric, we take \(a/r_h=0.5\). For the QCBH metric, we set \(\bar{\alpha}=0.5\). At the fixed
emission position, we calculate \(K_4(\Delta L)\) and \(\epsilon(\Delta L)\) for each metric. The results are
shown in Fig.~\ref{fig:ks_fixedx} and Fig.~\ref{fig:qcbh_fixedx}, where the upper panel shows the LG parameter
\(K_4(\Delta L)\), and the lower panel shows the corresponding effective-distance-interval
nonuniformity \(\epsilon(\Delta L)\). The specifical values of the parameters  are summarized in
Table~\ref{tab:fixedx_results}.

\begin{table*}[!htbp]
\centering
\caption{
Maximum LG parameter, corresponding proper interval, and
effective-interval nonuniformity at the peak position for different metrics and
propagation modes at fixed \(x=3\).
}
\label{tab:fixedx_results}
\begin{ruledtabular}
\begin{tabular}{ccccc}
Propagation mode & Metric & \(K_{4,\max}\) &
\(\Delta L_{\max}({\rm km})\) & \(\epsilon(K_{4,\max})\)\\
\hline
Radial & Schwarzschild
& 2.465749 & \(1.355236\times10^8\) & \(2.921537\times10^{-2}\)\\
Radial & KS
& 2.507520 & \(1.388439\times10^8\) & \(2.643264\times10^{-2}\)\\
Radial & QCBH
& 2.475083 & \(1.350035\times10^8\) & \(3.025284\times10^{-2}\)\\
\hline
Nonradial & Schwarzschild
& 2.395723 & \(1.390039\times10^8\) & \(4.163515\times10^{-2}\)\\
Nonradial & KS
& 2.525039 & \(1.418042\times10^8\) & \(3.707603\times10^{-2}\)\\
Nonradial & QCBH
& 2.432326 & \(1.276428\times10^8\) & \(4.375594\times10^{-2}\)\\
\end{tabular}
\end{ruledtabular}
\end{table*}

For the considered parameter, we see that both the KS and QCBH metrics give larger
\(K_{4,\max}\) values than the classical Schwarzschild result for radial and
nonradial propagation. This indicates that, given the emission position both the quantum-corrections can enhance the
observed quantumness in neutrino oscillations. However, this local enhancement does not prove that the classical
metric and the quantum-corrected metrics can be distinguished over the entire
emission-position range and for different quantum-correction parameters.

\subsection{Emission-Position Scan, Parameter Dependence, and Metric Discriminability}

To determine whether the apparent quantumness-enhancement still holds over a more general range of
emission positions, we further scan
the emission position \(x\) and compare the corresponding results for different
quantum-correction parameters.

\begin{figure*}[!t]
\centering
\safeincludegraphics[width=0.6\textwidth]{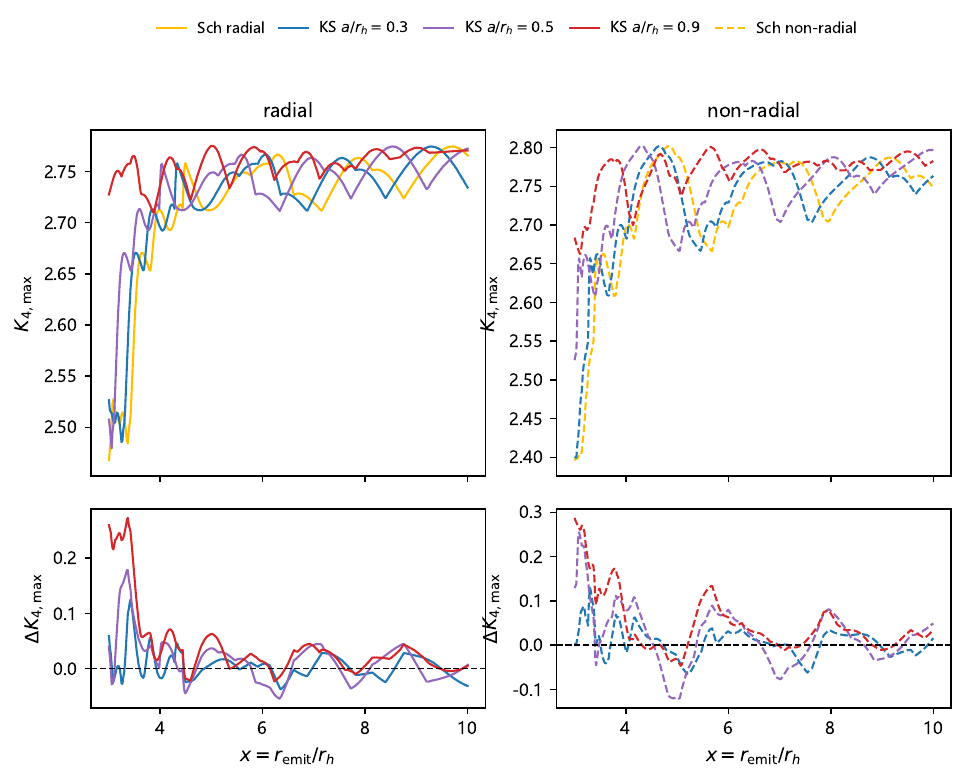}
\par\vspace{0.25em}
\vspace{0.3em}
\safeincludegraphics[width=0.58\textwidth]{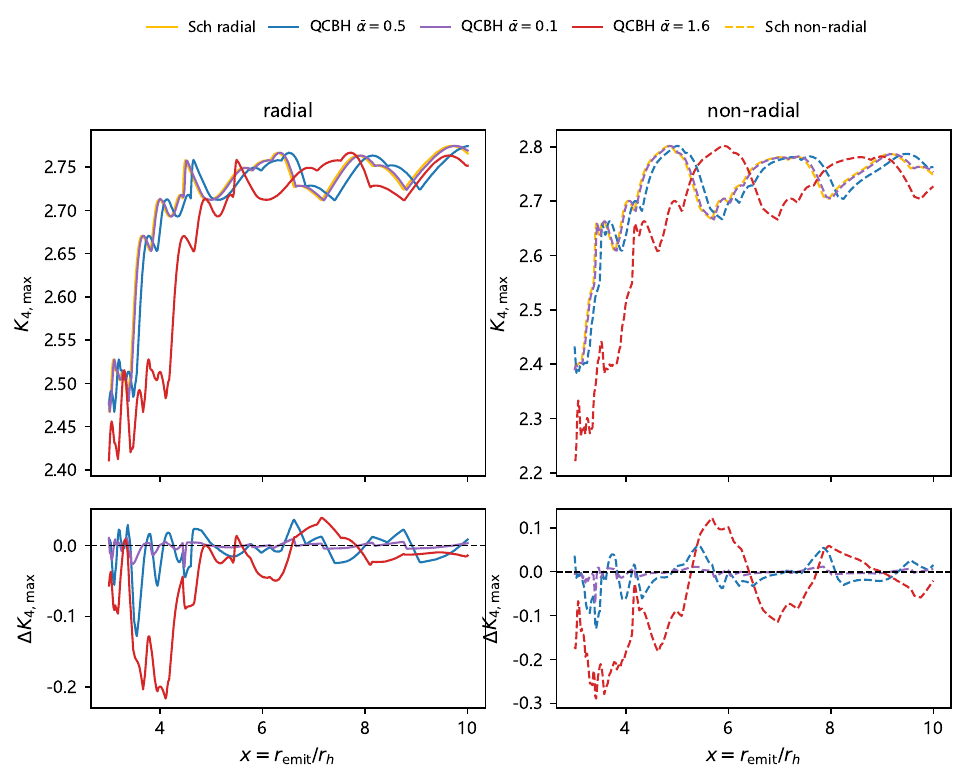}
\par\vspace{0.25em}
\caption{
Effect of the quantum-correction parameters on \(K_{4,\max}(x)\) at different
emission positions \(x=r_{\rm emit}/r_h\). The upper panel  shows the results for the KS
quantum-corrected metric with \(a/r_h=0.3,0.5,0.9\), and the lower shows the results
for the QCBH quantum-corrected metric with \(\bar{\alpha}=0.5,1.0,1.6\), with the
classical Schwarzschild curve shown as a reference in both cases. In each
subfigure, the upper row gives \(K_{4,\max}(x)\), while the lower row gives the
difference \(\Delta K_{4,\max}(x)\) relative to the classical Schwarzschild result.
The left and right panels correspond to radial(solid lines) and nonradial propagation (dashed lines),
respectively.
}
\label{fig:quantum_parameter_dependence}
\end{figure*}

For each \(x\), \(K_4(\Delta L;x)\) is computed within the
given \(\Delta L\) interval, and its maximum value
\(K_{4,\max}(x)\) is taken for comparison.
The results are shown in Fig.~\ref{fig:quantum_parameter_dependence}. The upper panel
gives \(K_{4,\max}(x)\) for the KS quantum-corrected metric with different values of
\(a/r_h\), together with its difference relative to the classical Schwarzschild
result. The lower gives the corresponding results for the QCBH quantum-corrected
metric with different values of \(\bar{\alpha}\).

It can be seen from the figure that the \(K_{4,\max}(x)\) curves obtained from the
quantum-corrected metrics are clearly different from the classical Schwarzschild
curve. For both types of quantum-corrected metrics, the difference becomes more
pronounced as the quantum-correction parameter increases. To quantify the
difference and examine whether it has a stable pattern, we define
\begin{equation}
\Delta K_{4,\max}(x)
=
K_{4,\max}^{\rm Q}(x)
-
K_{4,\max}^{\rm Sch}(x).
\end{equation}
Here, the superscript Q denotes the quantum-correction case. If
\(\Delta K_{4,\max}(x)>0\), the quantum-corrected metric enhances the apparent
 quantumness;
if \(\Delta K_{4,\max}(x)<0\), the maximal quantumness is suppressed.

Within the computational interval, \(\Delta K_{4,\max}(x)\) can be positive, negative, or even
close to zero at some positions. This shows that the main effect of the quantum-corrected metrics is
not a global enhancement or suppression of \(K_{4,\max}\), but a
moderate variation on the amplitude of the curve of \(K_{4,\max}(x)\) with respect to the
emission position and the quantum-correction parameter.
Therefore, although the classical and the quantum-corrected Schwarzschild metrics give different \(K_{4,\max}(x)\) curves,
the difference does not show a consistent pattern over different
emission positions.

\section{Comparison of the LG Parameter in Classical Kerr and Rotating KS Spacetimes }
\label{sec:Compa Kerr}

In the preceding section, we have seen that the apparent
\(K_{4,\max}(x)\) exhibits a pronounced response to the quantum-corrected Schwarzschild
metrics. However, the response does not show a stable enhancement or suppression for different emission positions and quantum-correction parameters.
Thus, the apparent  quantumness in neutrino oscillations with uniform proper-distance intervals cannot by itself serve as a universal indicator for
distinguishing the classical Schwarzschild metric from quantum-corrected metrics. To examine whether this observation applies to a rotating
black-hole, we further compares the \(K_{4,\max}\) parameter
in the classical Kerr and the rotating KS quantum-corrected metric.

The rotating KS metric is a quantum-corrected version of
the Kerr metric~\cite{Jusufi:2022spinningKS}.
Introducing
\begin{equation}
m_\beta(r)
=
M+\frac{1}{2}\left(r-\sqrt{r^2-\beta^2}\right),
\end{equation}
with \(\beta\) being the KS correction parameter,
the rotating KS metric can be written as
\begin{align}
ds^2_{\rm KS}
=&
-\left(1-\frac{2m_\beta(r)r}{\Sigma}\right)dt^2
-\frac{4a m_\beta(r)r\sin^2\theta}{\Sigma}\,dt\,d\phi
+\frac{\Sigma}{\Delta_{\rm KS}}dr^2
+\Sigma d\theta^2
\nonumber\\
&+
\left(
r^2+a^2+\frac{2a^2m_\beta(r)r\sin^2\theta}{\Sigma}
\right)\sin^2\theta\,d\phi^2 ,
\end{align}
where
\begin{equation}
\Sigma=r^2+a^2\cos^2\theta,
\end{equation}
and
\begin{equation}
\Delta_{\rm KS}(r)
=
r^2-2m_\beta(r)r+a^2
=
r\sqrt{r^2-\beta^2}
-2Mr
+a^2 .
\end{equation}
When \(\beta=0\), one has
\begin{equation}
m_\beta(r)\rightarrow M,
\qquad
\Delta_{\rm KS}(r)\rightarrow r^2-2Mr+a^2,
\end{equation}
and therefore the rotating KS metric reduces to the classical Kerr metric. On the other hand, when
\(a=0\), the rotating KS metric reduces to the quantum-corrected
Schwarzschild metric.

\begin{figure*}[!htbp]
\centering

\safeincludegraphics[width=0.93\textwidth]{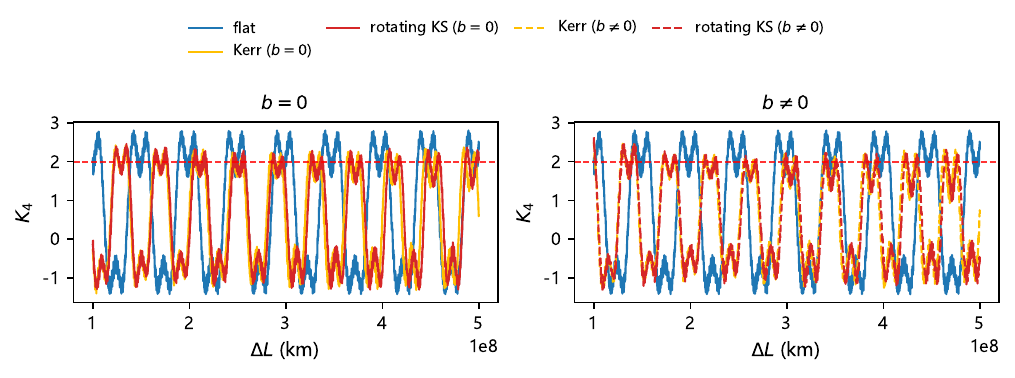}
\par\vspace{0.35em}
\vspace{0.5em}

\safeincludegraphics[width=0.93\textwidth]{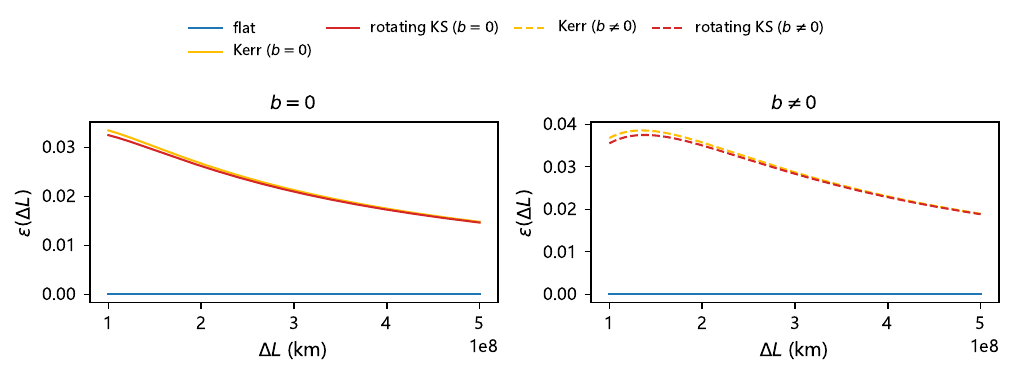}
\par\vspace{0.35em}

\caption{
The variation of \(K_4(\Delta L)\) and \(\epsilon(\Delta L)\) with \(\Delta L\) in
the classical and the quantum-corrected  Kerr metric, at the fixed emission position \(x=r_{\rm emit}/r_h=3\). The
quantum-correction parameter and the dimensionless spin parameter are fixed at
\(\beta/M=0.5\) and \(a/M=0.5\), respectively. The upper panel shows the
LG parameter \(K_4(\Delta L)\), and the lower panel shows the corresponding
effective-distance-interval nonuniformity \(\epsilon(\Delta L)\). In each panel, the left
and right subpanels correspond to propagation with \(b=0\) and \(b\neq0\),
respectively.
}
\label{fig:app_kerr_rotks_k4_epsilon}
\end{figure*}

\begin{figure*}[!htbp]
\centering
\safeincludegraphics[width=0.75\textwidth]{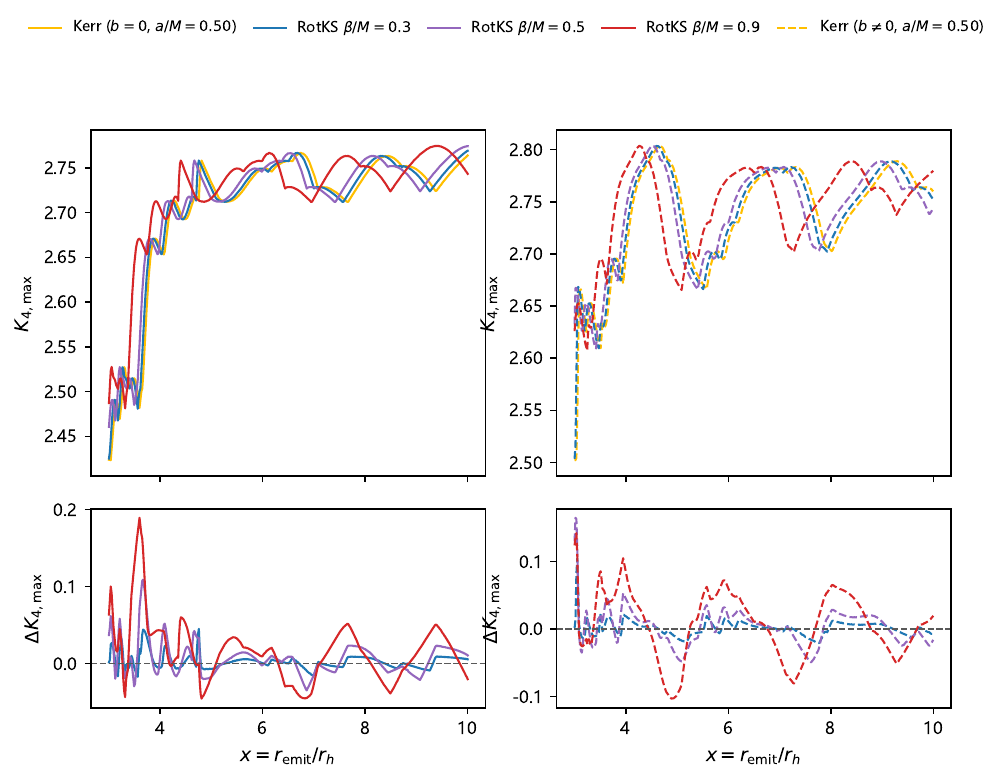}
\caption{
Effect of the rotating KS quantum-correction parameter on \(K_{4,\max}(x)\) at
different emission positions \(x=r_{\rm emit}/r_h\). The results are shown for the
rotating KS quantum-corrected metric with \(\beta/M=0.3,0.5,0.9\), with the
dimensionless spin parameter fixed at \(a/M=0.5\), and with the classical Kerr
curve shown as a reference. In each subfigure, the upper row shows
\(K_{4,\max}(x)\), while the lower row shows the difference
\(\Delta K_{4,\max}(x)\) relative to the classical Kerr result. The left and right
panels correspond to propagation with \(b=0\) (solid lines) and \(b\neq0\) (dashed
lines), respectively.
}
\label{fig:app_kerr_rotks_multi}
\end{figure*}

Employing the given metrics, we perform numerical calculations using the same parameters as those adopted in Sec.~III. The comparison
of \(K_4(\Delta L)\) and the effective-interval nonuniformity
\(\epsilon(\Delta L)\) at the fixed emission position \(x=3\), is shown in
Fig.~\ref{fig:app_kerr_rotks_k4_epsilon}. The comparison of \(K_{4,\max}(x)\) and
\(\Delta K_{4,\max}(x)\) for different emission positions and different
quantum-correction parameters is shown in
Fig.~\ref{fig:app_kerr_rotks_multi}.

The figures show that the effect of the rotating KS metric on
\(K_{4,\max}(x)\) is still not a simple global enhancement or global suppression.
As the quantum-correction parameter increases, \(\Delta K_{4,\max}(x)\) also
increases in a large portion of the \(x\)-range. However, within the computational
interval, \(\Delta K_{4,\max}(x)\) can be positive, negative, or close to zero at
some positions. Therefore, the quantum modification does not exhibit a consistent response
pattern over different emission positions.

\section{Conclusions}
\label{sec:Conclu}
Quantum coherence of particles in gravitational fields is an fundamental issue for understanding both the complete particle physics and the nature of spacetime.
For gravity-modulated  quantum coherence, how to separate the impact of intrinsic quantum properties of particles and the effects of spacetime is of particular significance.
To this end, we proposed an effective distance to characterize the essential gravitational effects on quantum phases. For neutrino propagations in Schwarzschild-, Kerr-type, and general asymptotically flat spacetimes,
we extended the oscillation probability in flat spacetime to that in curved spacetime  with a Hamiltonian formulation, which shows that the gravitational effects are completely determined by the effective distance.
Employing effective distances, we proved that one can always map a neutrino oscillation phase in  asymptotically flat  spacetime onto the one in flat spacetime, which can be viewed as  a nontrivial realization of equivalence principle.
Based on the observation, we resolved the paradox that the gravitational effects on quantum coherence of neutrinos is measure-dependent. For LG parameters, employing
the measurement scheme where uniform effective-distance intervals are taken, we showed that the maximum is independent from spacetime backgrounds, which takes the unique value determined by
the properties of neutrino itself.  We demonstrated that this observation also applies to the gravitational effects on general quantum information quantities in terms of neutrino-oscillation phases.

To demonstrate the independence of quantum coherence from spacetime, we furthermore examined the proposal  employing the LG parameter to identify quantum-corrected models of spacetimes. For Schwarzschild and Kerr metrics, the effects of considered quantum corrections on the observed coherence do not show a consistent pattern for different emission positions of neutrinos and various quantum parameters.
The results show that one cannot discriminate the classical and quantum-corrected gravities from quantum coherence in neutrino oscillations. This observation is consistent with the conclusion that there exists no certain distinction
between quantum entanglement mediated by the Newtonian gravity and entanglement mediated by gravitons~\cite{Danielson:2021egj}. On the contrary,  Ref.~\cite{Petruzziello:2023xhb} argued that the Bell's inequality can be used to  discriminate the classical, the quantum-corrected, and the quantum gravity. However, their observation is mainly based on the different oscillation behaviors of the Bell parameters due to the different metrics, yet no pattern is identifiable in their oscillation curves to show a quantum feature of spacetime.

Taken together, our work provides a useful notion to identify the essential gravitational effects on quantum coherence of relativistic particles and  show a simple approach to resolving the reported paradoxes on  Leggett-Garg inequality and
quantum information quantities on neutrinos.

\acknowledgments
We thank Tian-Hao Wu and Ze-Wen Li for their helpful discussions.
This work was supported by the National Natural Science Foundation of China under grant No. 12565015 and the Natural Science Foundation of Guangxi under grant No. 2026GXNSFAA00640923.

\appendix
\section{Hamilton--Jacobi Derivation of  Covariant Phases for Neutrinos}
\label{app:covariant-phase}

\renewcommand{\theequation}{A\arabic{equation}}
\setcounter{equation}{0}

Considering a particle of mass \(m\) propagating in curved spacetime, the action
\(S\) and the canonical four-momentum \(p_\mu\) satisfy
\(p_\mu=\partial_\mu S\). Hence, along the propagation trajectory,
\(dS=p_\mu dx^\mu\). For the metric signature \((-+++)\), the
Hamilton--Jacobi equation is
\begin{equation}
g^{\mu\nu}
\partial_\mu S
\partial_\nu S
+
m^2
=
0.
\label{eq:A_HJ}
\end{equation}
In the relativistic limit \(m^2/E^2\ll1\), the action can be expanded as~\cite{Wudka:2000rf,Visinelli:2014xsa}
\begin{equation}
S
=
S_0
+
m^2S_1
+
O(m^4).
\label{eq:A_action_expansion}
\end{equation}
Substituting this expansion into the Hamilton--Jacobi equation and comparing
the zeroth-order and \(O(m^2)\) terms yields
\begin{equation}
g^{\mu\nu}
\partial_\mu S_0
\partial_\nu S_0
=
0,
\label{eq:A_null_HJ}
\end{equation}
and
\begin{equation}
2g^{\mu\nu}
\partial_\mu S_0
\partial_\nu S_1
+
1
=
0.
\label{eq:A_first_order_HJ}
\end{equation}

The four-momentum associated with the common null reference trajectory is
defined by
\begin{equation}
k_\mu
=
\partial_\mu S_0,
\qquad
k^\mu
=
\frac{dx^\mu}{d\lambda},
\label{eq:A_null_momentum}
\end{equation}
where \(\lambda\) is an affine parameter. It follows that
\begin{equation}
\frac{dS_1}{d\lambda}
=
-\frac{1}{2}.
\label{eq:A_S1_derivative}
\end{equation}
The mass-dependent increment of the action is therefore\cite{Wudka:2000rf}
\begin{equation}
dS^{(m)}
=
m^2\,dS_1
=
-\frac{m^2}{2}\,d\lambda.
\label{eq:A_mass_action}
\end{equation}

In the WKB approximation, the wave function of the \(i\)th mass eigenstate can
be written as \(\psi_i\propto e^{iS_i}\). If the propagation phase is defined
through \(\psi_i\propto e^{-i\Phi_i}\), then \(\Phi_i=-S_i\), and hence
\(d\Phi_i=-dS_i^{(m)}\). Setting the mass in
Eq.~\eqref{eq:A_mass_action} to \(m_i\), one obtains
\begin{equation}
d\Phi_i
=
\frac{m_i^2}{2}\,d\lambda.
\label{eq:A_phase_i}
\end{equation}
The corresponding oscillation phase difference is
\begin{equation}
d\Phi_{ij}
=
\frac{\Delta m_{ij}^2}{2}\,d\lambda.
\label{eq:A_phase_ij_affine}
\end{equation}

For a local observer with four-velocity \(u^\mu\), the locally
measured energy is defined as
\begin{equation}
E_{\rm loc}
=
-k_\mu u^\mu.
\label{eq:A_local_energy}
\end{equation}
Along the null reference trajectory,
\begin{equation}
dx^\mu
=
k^\mu d\lambda.
\label{eq:A_null_displacement}
\end{equation}
The corresponding local time interval is therefore
\begin{equation}
dT_{\rm loc}
=
-u_\mu dx^\mu
=
-u_\mu k^\mu d\lambda
=
E_{\rm loc}\,d\lambda.
\label{eq:A_local_time}
\end{equation}
In the local orthonormal frame, the null condition gives
\begin{equation}
0
=
ds^2
=
-dT_{\rm loc}^2
+
dL_{\rm loc}^2.
\label{eq:A_local_null_condition}
\end{equation}
Taking the positive root yields
\begin{equation}
dL_{\rm loc}
=
dT_{\rm loc}.
\label{eq:A_local_length_time}
\end{equation}
Combining the above relations gives
\begin{equation}
dL_{\rm loc}
=
E_{\rm loc}\,d\lambda,
\qquad
d\lambda
=
\frac{dL_{\rm loc}}{E_{\rm loc}}.
\label{eq:A_affine_local_distance}
\end{equation}
The covariant oscillation phase can thus be expressed as
\begin{equation}
d\Phi_{ij}
=
\frac{\Delta m_{ij}^2}{2E_{\rm loc}}
\,dL_{\rm loc}.
\label{eq:A_local_phase}
\end{equation}
Integrating along the propagation path finally gives
\begin{equation}
\Phi_{ij}(L_{\rm loc})
=
\frac{\Delta m_{ij}^2}{2}
\int_0^{L}
\frac{dL_{\rm loc}'}
{E_{\rm loc}\!\left[x(L_{\rm loc}')\right]}.
\label{eq:A_integrated_phase}
\end{equation}
Here, \(x(L_{\rm loc}')\) collectively denotes the position along the trajectory
and may include \(r(L_{\rm loc}')\), \(\theta(L_{\rm loc}')\), and any other
coordinates on which the locally measured energy depends.

\bibliographystyle{apsrev4-1}
\bibliography{ref}

\end{document}